\documentclass[paper]{ieice}
\usepackage[pdftex]{graphicx,xcolor}
\usepackage{color, soul}
\usepackage[fleqn]{amsmath}
\usepackage{newtxtext}
\usepackage[varg]{newtxmath}

\usepackage{tabularx}
\usepackage{booktabs}
\usepackage{collcell}
\usepackage{hhline}
\usepackage{multirow}

\newcolumntype{L}[1]{>{\raggedright\let\newline\\\arraybackslash\hspace{0pt}}m{#1}}
\newcolumntype{C}[1]{>{\centering\let\newline\\\arraybackslash\hspace{0pt}}m{#1}}
\newcolumntype{R}[1]{>{\raggedleft\let\newline\\\arraybackslash\hspace{0pt}}m{#1}}

\newcommand{\dtoprule}{\specialrule{1pt}{0pt}{0.4pt}%
            \specialrule{0.3pt}{0pt}{\belowrulesep}%
            }
\newcommand{\dbottomrule}{\specialrule{0.3pt}{0pt}{0.4pt}%
            \specialrule{1pt}{0pt}{\belowrulesep}%
            }

\field{}
\title{Speech Paralinguistic Approach for Detecting Dementia Using Gated Convolutional Neural Network}
\authorlist{%
\authorentry{Mariana Rodrigues Makiuchi}{}{titech}\MembershipNumber{}
\authorentry{Tifani Warnita}{}{titech}\MembershipNumber{}
\authorentry{Nakamasa Inoue}{}{titech}\MembershipNumber{}
\authorentry{Koichi Shinoda}{}{titech}\MembershipNumber{}
\authorentry{Michitaka Yoshimura}{}{keio}\MembershipNumber{}
\authorentry{Momoko Kitazawa}{}{keio}\MembershipNumber{}
\authorentry{Kei Funaki}{}{keio}\MembershipNumber{}
\authorentry{Yoko Eguchi}{}{keio}\MembershipNumber{}
\authorentry{Taishiro Kishimoto}{}{keio}\MembershipNumber{}
}
\affiliate[titech]{The authors are with Department of Computer Science, School of Computing, Tokyo Institute of Technology, Tokyo, Japan}
\affiliate[keio]{The authors are with Department of Neuropsychiatry, Keio University School of Medicine, Tokyo, Japan}
\begin{document}
\maketitle
\begin{summary}

%!TEX root =  ../elsarticle.tex
%%abstract%%

We propose a non-invasive and cost-effective method to automatically detect dementia by utilizing solely speech audio data. We extract paralinguistic features for a short speech segment and use Gated Convolutional Neural Networks (GCNN) to classify it into dementia or healthy. We evaluate our method on the Pitt Corpus and on our own dataset, the PROMPT Database. Our method yields the accuracy of 73.1\% on the Pitt Corpus using an average of 114 seconds of speech data. In the PROMPT Database, our method yields the accuracy of 74.7\% using 4 seconds of speech data and it improves to 80.8\% when we use all the patient's speech data. Furthermore, we evaluate our method on a three-class classification problem in which we included the Mild Cognitive Impairment (MCI) class and achieved the accuracy of 60.6\% with 40 seconds of speech data.

\end{summary}


\section{Introduction}
\label{introduction}
%\begin{itemize}
% \item Definition of dementia
% \item Importance of automatic diagnosis
% \item Challenges in the dementia diagnosis (try to answer why a diagnosis that requires only 4 seconds of data would be useful)
% \item Brief introduction to the current dementia detection approaches and to their drawbacks
% \item Brief outline of our method, highlighting its benefits (i.e., novelty, efficiency, significance) when compared to existing approaches
% \item Highlight the efficiency of GCNNs when there is data insufficiency (?)
%\end{itemize}

Dementia is an umbrella term for a group of medical signs and symptoms associated with the cognitive-related deficiency due to damage in neurons \cite{alzheimer20182018}. Types of dementia include Alzheimer's disease (AD), vascular dementia, dementia with lewy body (DLB) and frontotemporal lobar degeneration (FTLD). Dementia have various characteristics representing cognitive dysfunction such as poor narrative memory when recalling experiences \cite{prud2011extraction} as well as difficulties in making plans, solving problems, and completing daily tasks \cite{alzheimer20182018}.

The increasing number of people living with dementia has gained a lot of attention. AD, which takes the biggest proportion of dementia, has become the 6\textsuperscript{th} leading cause of death in the United States of America \cite{alz2017}. Moreover, according to the World Health Organization \cite{dementia2017who}, in 2015, dementia affected 47 million people worldwide and it is estimated that, by 2050, this number will be nearly triplicated.

% Sebenernya belum kasih referensi buat yang approaches ini 
Unfortunately, there is no clear protocol on how to detect dementia in an accurate and effective manner \cite{alz2017}. The most common approach is to perform various clinical assessments of the patients such as examining their medical history, conducting cognitive tests (e.g., memory tasks, executive function tasks, picture description tasks, naming tasks), assessing their mood and mental status, as well as performing brain imaging; i.e., computerized tomography (CT), magnetic resonance imaging (MRI), single-photon emission computed tomography (SPECT), positron emission tomography (PET), and blood/cerebrospinal fluid testing.

The careful diagnosing process can be invasive, time-consuming and costly. Early cognitive deficiency treatments can help patients preserve their cognitive functions \cite{martono2000buku} as some causes of dementia are remediable in early stages \cite{tripathi2009reversible}. Faster and more cost-effective dementia detection approaches have been strongly demanded.

%The detection of MCI has been extensively studied \cite{tuokko2020mild} and it is considered extremely important since it is estimated that an annual average of 10\% to 15\% of people with MCI might progress to dementia \cite{arevalo2015mini}. Early cognitive deficiency treatments can help patients preserve their cognitive functions \cite{martono2000buku} as some causes of dementia are remediable in early stages \cite{tripathi2009reversible}.

% Unfortunately, the amount of MCI data is small and its study is suffered from the data sparseness problem.

Most approaches for the automatic dementia detection relied on linguistic information \cite{zimmerer2016formulaic,fraser2016linguistic,orimaye2017predicting,wankerl2017n,mirheidari2018detecting} since cognitive dysfunctions in patients typically appear as linguistic impairments. While these methods are effective, their major drawback is the requirement of transcriptions of patient's speech. Manual transcription is costly, and Automatic Speech Recognition (ASR) is often erroneous.

% luz2018method : dementia detection based on speech characteristics (vocalization events) ("Features analysed included speech rate, turn-taking patterns and other speech parameters.  Despite relyingsolely on content-free features, our method obtains overall accuracy of 86.5%, a result comparable to those of state-of-the-art methodsthat employ more complex lexical, syntactic and semantic features"). Their database is called Carolina Conversations  Collections (CCC) and they only consider data from 38 people.

% mirheidari2019dementia: also uses ASR and apply feature extractors in order to get lexical, acoustic and Conversation Analysis (CA) inspired features. It combines similar acoustic features as luz2018method and other several features. They collected their own database, in which they achieve 90% accuracy.

In order to address this issue, we propose a dementia detection method that relies on speech audio data only. Moreover, since geriatric clinical assessment presents several challenges \cite{singh2016assessment} and patients may feel fatigued by it, this work focuses on using less patient speech data as possible to make the diagnosis in real time and physically less demanding for the patients.

We employ Gated Convolutional Neural Networks (GCNN) in order to capture the temporal pattern in the extracted features. We chose the GCNN architecture due to its superior performance in several tasks \cite{dauphin2016language,oord2016wavenet}, including tasks with limited amount of data \cite{kang2017gated}. As an extension of our previous work \cite{Warnita2018}, we evaluate our method on two datasets, the DementiaBank Pitt Corpus (English) and the PROMPT Database (Japanese) collected by our own.

We further explore the detection of Mild Cognitive Impairment (MCI) patients. MCI is the stage of cognitive impairment between the expected cognitive decline of normal ageing and early dementia \cite{petersen2004mild}. MCI is characterized by a cognitive decline that is greater than the age-related expectation, but that cannot be defined as dementia yet \cite{petersen1999mild}.

The experimental results demonstrate the effectiveness of our method in terms of accuracy, cost, and time required for each prediction.

% Our contributions include the following; First, we present a deep learning method for detecting dementia by using only speech data; Second, we introduce a hierarchical structure of binary classifiers for handling the imbalance case; Third, we evaluate our method on two datasets with different languages, which are English and Japanese. 

\section{Related Works}
\label{related_works}
\subsection{Dementia Assessment}
\label{sec:wk_dementia_assessment}

%\begin{itemize}
%\item How to assess dementia (and MCI. Do not emphasize MCI too much, but state the difficulty in the diagnosis)
%\item The metrics used to assess the severity of them
%\end{itemize}

Various evaluation methods have been defined to diagnose dementia in people. In the medical field, commonly used approaches are the Clinical Dementia Rating (CDR) \cite{morris1993clinical}, the Clock Drawing Test (CDT) \cite{sunderland1989clock}, the Neuropsychiatric Inventory (NPI) \cite{cummings1994neuropsychiatric} and the Mini-Mental State Examination (MMSE) \cite{folstein1975mini}. CDR uses an interview protocol to assess dementia severity as mild or severe while, in CDT, the patients are asked to draw a clock with a specific time to get a score, which can aid the diagnosis of neurological disorders. The NPI test assess the disruptions of several behavioural functioning.

The MMSE is an extensively-used screening test that quantifies patients' cognitive function as the total score of a series of questions and problems \cite{pangman2000examination}. The test itself is designed to aid the dementia diagnosis. The MMSE is used in this work and, based on \cite{saxton2009computer,kaufer2008cognitive,tombaugh1992mini,arevalo2015mini}, we define the score ranges of 0--23, 24--26 and 27--30 to respectively represent dementia, MCI and healthy categories. While most of the current medical research works present similar cut-off points to represent these classes, the definition of those score ranges has not been standardized \cite{onwuekwe2012assessment,o2008detecting,kochhann2010mini}. In addition, there is not a consensus about the definition and diagnosis of MCI, since its symptoms are vast and subtle \cite{arevalo2015mini}.

%In order to evaluate the efficiency of automatic dementia approaches, most works adopt the accuracy score as their performance measure \cite{fraser2016linguistic,wankerl2017n,khodabakhsh2015evaluation,sadeghian2017speech}. 
%Additionally, in the medical field, the Cohen's Kappa score is often used to evaluate a classifier's performance \cite{pezzotti2008accuracy,wang2003comparison}. The Kappa score measures the observed accuracy from the classifier's outcome, normalized by the expected accuracy from random chance.

\subsection{Features}
\label{sec:wk_features}

%Types of data used for the dementia detection and the features that can be extracted from them.

Several types of features can be used to identify dementia. In this section, features extracted from the patient's brain images and from the their speech's linguistic content will be presented as well as acoustic features obtained from their speech, which are the features used in this work.

\subsubsection{Image}
\label{sec:wk_image}

%Feature extraction from images.

Structural brain images from Magnetic Resonance Imaging (MRI) can be used to identify AD patients \cite{farhan2014ensemble}. Brain imaging plays an important role in neurodegenerative disorder detection because it provides useful information regarding anatomical changes in patients' brain \cite{narayanan2016can}. The combination of MRI and fluorodeoxyglucose-positron emission tomography (FDG-PET) images was used to identify MCI patients who would further progress to dementia \cite{lu2018multimodal}. Despite their effectiveness, medical image acquisition is costly and not easily accessible.

\subsubsection{Linguistic data}
\label{sec:wk_linguistic}

%Feature extraction from linguistic data (includes the ASR-based approaches).

Language deficiency becomes a prominent and perceivable symptom of dementia patients. Several syntactic, lexical and \textit{n}-gram features were used for detecting AD on the DementiaBank Pitt Corpus \cite{zimmerer2016formulaic,orimaye2017predicting}. \cite{wankerl2017n} used \textit{n}-gram and MMSE score correlation analysis. More recently, \cite{chen2019attention} defined a hybrid RNN-CNN architecture with an attention mechanism to detect AD from Pitt Corpus' transcriptions' textual embeddings. \cite{fritsch2019automatic} proposed a LSTM-based neural network language model whose prediction is calculated from their model's perplexity.

Several other works have studied the combination of linguistic and acoustic features. \cite{mirheidari2019dementia} combined features inspired in the conversation analysis of clinical interviews, lexical information extracted with an ASR and acoustic features. They further input these features to a support vector machine (SVM) to classify the patients of their own dataset into dementia or functional memory disorder. \cite{gosztolya2016detecting} and \cite{toth2015automatic} extracted phonetic-based features with an ASR in order to detect MCI patients from their speech. \cite{fraser2016linguistic} fused transcription-based linguistic features with acoustic features such as Mel-frequency cepstral coefficients (MFCC). \cite{sadeghian2017speech} utilized the combination of speech duration, pause-related features, pitch-related features and other prosodic features, as well as linguistic features acquired from a customized ASR adapted for dementia patients.

Even though those approaches have shown good results, most of them are limited by the availability of transcription data and/or ASR, which often has poor performance due to the degraded speech intelligibility of the patients.

\subsection{Speech}
\label{sec:wk_speech}

Several works proposed the dementia identification from ASR-independent speech features. Features such as silence ratio were found to be more meaningful than other linguistic features when applied to a SVM classifier \cite{khodabakhsh2015evaluation}. Moreover, the usage of acoustic and context-free linguistic features to classify patients showed promising results on the Carolina Conversations Collections dataset \cite{luz2018method}.

Besides having problems with language deficits, people with AD, specially in the early stages of the disease, might become apathetic and have a tendency to get depressed \cite{alzheimer20182018}. People with AD usually suffer from prosodic impairment due to which they will find difficulties in expressing their emotions \cite{tosto2011prosodic}. Those signs suggest the presence of paralinguistic cues in the speech of people who suffer from this cognitive dysfunction.

OpenSMILE \cite{eyben2013recent} is a commonly used tool for feature extraction in speech tasks \cite{schuller2009interspeech,schuller2010interspeech}. It describes a series of default feature sets, such as the INTERSPEECH 2010 Paralinguistic Challenge Feature Set (IS10) \cite{schuller2010interspeech}, which is used in this work. The IS10 defines 76 Low-Level Descriptors (LLD) features for each time frame. In this work, we define a time frame as 25ms, sampled at a rate of 10ms. Those LLD are a combination of several speech descriptors that were independently used in previous works, such as pitch, voicing probability \cite{khodabakhsh2015evaluation} and MFCC \cite{fraser2016linguistic}. When compared to other feature sets defined in OpenSMILE, the IS10 yielded the best result in the AD detection task \cite{Warnita2018}.

\subsection{Classifiers}
\label{sec:wk_classifiers}

%Methods used to classify the features extracted from the patient's data as dementia or healthy.

Support Vector Machine (SVM) classifiers \cite{boser1992training} were widely used as the baseline method of several paralinguistic tasks, such as emotion recognition \cite{schuller2009interspeech} and age-gender classification \cite{schuller2010interspeech}. While training the SVM, the Sequential Minimal Optimization (SMO) \cite{platt1998sequential} algorithm is used. The SMO solves the Quadratic Programming (QP) optimization in the SVM by dividing the QP into the smallest possible QP sub-problems, allowing the SMO to handle large amounts of data.

Recently, deep learning-based approaches have become extremely popular due to their success in a wide range of tasks. \cite{huang2014speech} applied Convolutional Neural Networks (CNN) \cite{lecun1989backpropagation} to speech emotion recognition. For the same task, a Recurrent Neural Network (RNN) \cite{hopfield1982neural} was added on top of CNN layers \cite{keren2016convolutional} to capture the speech's dynamic features.

RNNs can accommodate the temporal pattern change, but they require a long training time \cite{peddinti2015time} and a large amount of training data. On the other hand, CNNs need little training data compared to other existing networks \cite{bengio2009learning} due to their reduced number of connection weights. Moreover, even without any explicit sequential mechanism, CNNs are still able to model the temporal context in the data by means of their convolution operations \cite{peddinti2015time}.

There have been various studies that incorporated gating mechanisms to convolution layers achieving state-of-the-art performance on tasks such as conditional image modelling \cite{oord2016conditional}, language modelling \cite{dauphin2016language}, speech synthesis \cite{oord2016wavenet} and generative image inpainting \cite{yu2018free}. When applied to RNNs, such as Long Short-Term Memories (LSTM) \cite{hochreiter1997long}, gating mechanisms were shown to be effective in handling the long-term dependencies problem. Gating mechanisms in CNNs can be used to manage the information flow as well as to mitigate the vanishing gradient problem \cite{dauphin2016language}. Moreover, it was shown that the combination of gating mechanisms and CNNs can achieve superior performance in tasks with limited amount of data, such as speech recognition for low-resource languages \cite{kang2017gated} and speech keyword spotting \cite{coucke2019efficient}. Therefore, inspired by these advantages and the effectiveness of the combination of CNNs and gating mechanisms applied to different tasks, we hypothesize that the automatic dementia detection can also benefit from this architecture.

\section{Gated Convolutional Neural Network}
\label{method}
%This is the only section for the methodology. It should contain a complete explanation about the gated convolutional neural network operation. Highlight the usage of GCNNs for cases with data insufficiency.

A Gated Convolutional Neural Network (GCNN) consists of convolution layers and gating mechanisms. In our case, each convolution layer is expected to extract the salient information from the combined LLD features for every short period of time. Thus, the temporal pattern change will be encapsulated within the combination of several extracted patches of features.

\begin{figure}[tb]
  \centering
  \includegraphics[width=80mm]{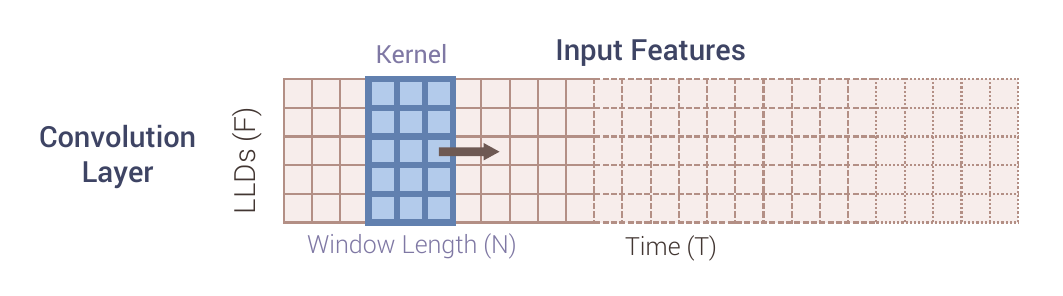}
  % \vspace{-1em}
  \caption{A convolution layer over LLD features extracted with the openSMILE toolkit.}
  \label{fig:conv_lld}
  % \vspace{-1.5em}
\end{figure}

The convolution operation ``slides'' a kernel \(k\) over the input features in order to extract their prominent cues. In our study, since we want to model the correlation between all the LLD features, captured at each time frame, we use the one-dimensional (1D) CNN, hence each kernel slides only in the time axis, as represented in Figure~\ref{fig:conv_lld}. This network is also referred to as Time-Delay Neural Network (TDNN) \cite{waibel1990phoneme}.

\begin{figure}[tb]
  \centering
  \includegraphics[width=80mm,scale=0.5]{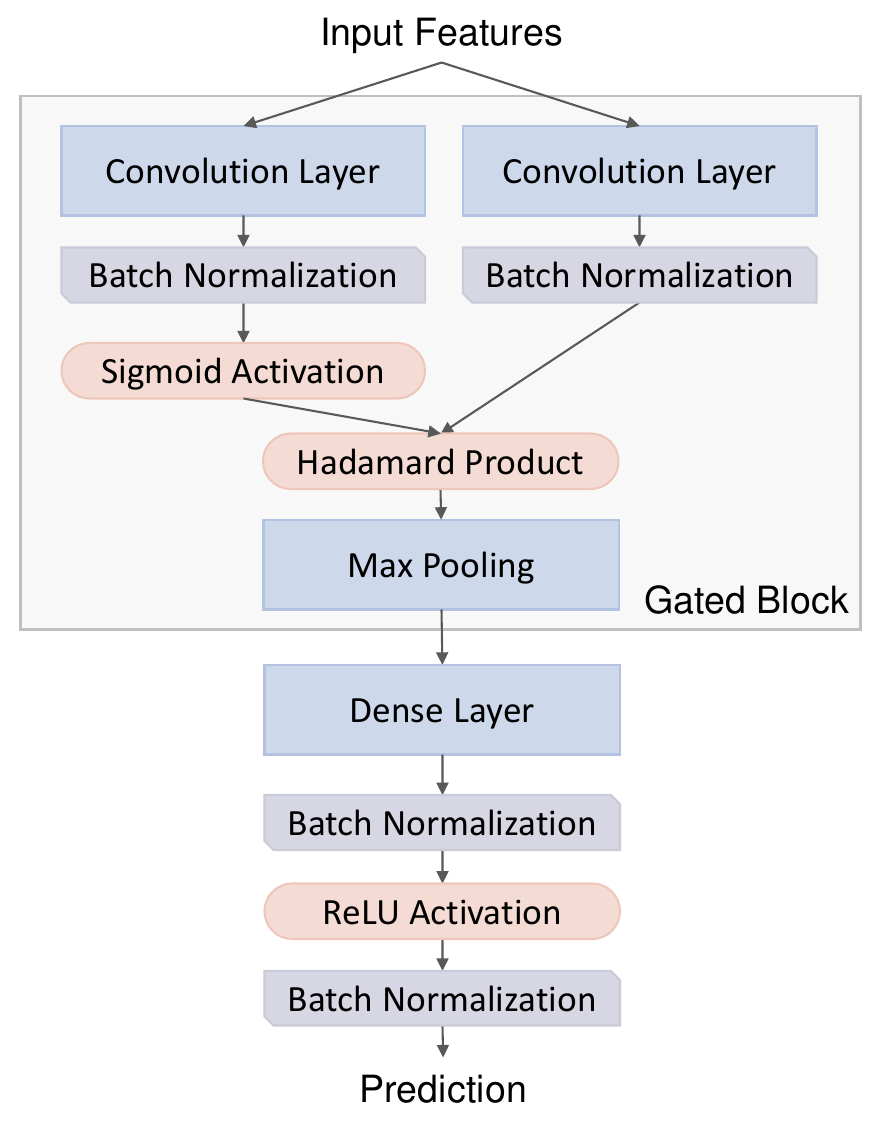}
  \vspace{-1em}
  \caption{A GCNN with one gated block. A deeper network can be made by stacking gated blocks.}
  \label{fig:framework_ov}
  \vspace{-1.5em}
\end{figure}

The gating mechanism applies this convolution operation to the input in two different paths, as shown in Figure~\ref{fig:framework_ov}. The gate in this network controls the information flow between succeeding layers, hence preventing the vanishing gradient problem.

%The gated linear unit (GLU) is chosen because it outperformed the gated tanh unit (GTU) \citep{dauphin2016language}.

% GCNN Structure

Following the Gated Linear Unit (GLU) architecture proposed by \cite{dauphin2016language}, we feed a speech feature matrix \(X \in \mathbb{R}^ {F \times T}\) into our network, in which \(F\) and \(T\) are the dimension of the LLD features and the number of time frames, respectively. We further convolve these input features with a kernel of dimension \(F \times N\), in which \(N\) is the length of the kernel in the time axis. At each convolution operation between this kernel and the input, a scalar output is produced. The result of the gated convolution operation before the max pooling for the kernel \(k\) at position \(i\) is given by

\begin{eqnarray}
\label{eq:gcnn}
y_{k,i} &=& \left(\sum\limits_{f = 1}^{F} \sum\limits_{n = 1}^{N} v_{(k)_{f,n}}x_{f,i-n} + e_{(k)}\right) \cdot \nonumber \\
& & \mbox{}g\left(\sum\limits_{f = 1}^{F} \sum\limits_{n = 1}^{N} w_{(k)_{f,n}}x_{f,i-n} + b_{(k)}\right),
\end{eqnarray}

\noindent
in which \(v_{(k)_{c,d}}\) is the element of kernel \(k\)'s matrix \(V\) at position \((c,d)\) and \(w_{(k)_{c,d}}\) is the element of kernel \(k\)'s matrix \(W\) at position \((c,d)\). \(V \in \mathbb{R} ^ {F \times N}\) and \(e \in \mathbb{R}\) are the linear gate kernel weight matrix and bias respectively (i.e., they represent the convolution operation in the right stream of the gated block in Figure~\ref{fig:framework_ov}), \(W \in \mathbb{R} ^ {F \times N}\) and \(b \in \mathbb{R}\) are the respective weight matrix and the bias of the convolutional operation in the left stream of Figure~\ref{fig:framework_ov} and \(g\) is the sigmoid function.

In Equation~(\ref{eq:gcnn}), both summations enclosed by parenthesis represent a convolution operation that results in one scalar. In addition, the term to which the sigmoid function \(g\) is applied is the gate operation that controls the linear gate output.

The resulting \(Y\) matrix formed by the elements \(y_{k,i}\) will have the dimensions \({K \times M}\), in which \(K\) is the number of kernels and \(M\) is the output segment length. After the convolution operation, the matrix \(Y\) has its length halved in the time-axis by the max-pooling layer \cite{zhou1988computation} to get its most significant information and reduce its dimensionality.

Figure~\ref{fig:framework_ov} shows our GCNN with a single gated block, which is formed by two convolution layers and one max-pooling layer. A deeper GCNN would consist of multiple gated blocks.

The output of the network's last gated block, \(Y' \in \mathbb{R} ^ {K' \times M'}\), with \(K'\) as the number of kernels of the last gated block and \(M'\) as the final output segment length, is then flattened into one feature vector \(Z \in \mathbb{R} ^ O\), in which \(O = K'M'\). This vector is input to a fully-connected (dense) layer with the ReLU activation function. We also apply batch normalization \cite{ioffe2015batch} at the end of each dense and convolution layer.

\section{Experiments}
\label{experiments}
%All the experiments, results, analysis and discussion.
%Clearly explain how to provide the data (VAD, units, etc)

\subsection{Evaluation Metrics}
\label{sec:exp_measures}

%Evaluation metrics used in this work. How to get the session-level and the utterance-level predictions.

We use classification accuracy as the main evaluation metric in our experiments. This reflects previous works on the Pitt Corpus dataset \cite{fraser2016linguistic,wankerl2017n} and on the other related datasets \cite{khodabakhsh2015evaluation,sadeghian2017speech}. We compute the accuracy averaged over the 10-fold cross-validation results. At each fold, we partition the dataset in 10 subsets, from which we select one for testing and the remaining for training. We design these subsets so that no subject's data appears in both training and testing.

\subsection{Configuration}
\label{sec:exp_configuration}

%Size of convolution kernels, number of layers, framework used, normalizations and optimizers, etc.

We split the patient's speech of each interview session into several segments of length \(L\), zero-padding the speech segments that are shorter than \(L\). In this work, a speech segment is defined as a short slice of the patient's speech. In order to obtain the speech segments, we first concatenate all the patient's speech utterances in an interview session. Then, we segment this combined patient's speech into consecutive and non-overlapping speech segments, which are applied as input to our model. On the Pitt Corpus, the patient's speech is extracted by using the speaker turns information. On the PROMPT Database we separate the patient and doctor speech by applying the Cross-Channel Spectral Subtraction method \cite{nasu2011cross} followed by a Voice Activity Detection (VAD) approach. We classify each speech segment using our Gated Convolutional Neural Network architecture, and, after aggregating the scores from multiple segments, we conduct a majority voting to determine the session-level dementia classification.

In our binary GCNN, we consistently use the window length \(N = 2\) and the kernel size \(K = 64\) in every convolution operation of each gated block. We have tested our model with 6, 8 and 10 stacked gated blocks. The dense layer after the last gated block has 256 hidden neurons. We apply 0.5 dropout before the output layer for regularization. The output layer consists of one neuron with a sigmoid function. We trained the network using each segment's IS10 LLD features and their corresponding binary label (i.e., healthy or AD) on a 10-fold cross-validation scheme.

We used binary cross-entropy as the loss function and the Adam \cite{kingma2014adam} optimizer with learning rate equal to $10^{-3}$ and exponential decay rate respectively defined as $0.9$ and $0.999$ for the first and second moment estimates. A batch size equal to $32$ was consistently used over all the experiments and the input \(X \in \mathbb{R} ^ {F \times T}\) is composed of $76$ LLD features \(F\) per time frame and $397$ time frames \(T\). 

\subsection{Datasets}

In this work, we use two datasets containing the speech of people with and without dementia: the Pitt Corpus \cite{becker1994natural}, in which the subjects speak English, and the PROMPT Database, in which the subjects speak Japanese.

\subsubsection{Pitt Corpus}
\label{sec:db_pitt}

%Present the Pitt Corpus (cover the same points as in the PROMPT database section)

The Pitt Corpus, a part of the DementiaBank, contains speech data and the corresponding transcription information of healthy people (Control group) and of people with Alzheimer's disease (AD group) speaking in English. The audio files in this dataset contain speech from clinicians and patients. We apply three constraints to select data from this dataset.

First, we only consider the data drawn from the picture description task. This task is considered an approximation of real-life spontaneous conversations \cite{giles1996performance}, in which the subjects are asked to describe the Cookie Theft Picture of the Boston Diagnostic Aphasia Examination \cite{kaplan1983boston}.

Second, from the AD group, we select the sessions that correspond to patients with a diagnosis of either AD or probable AD. There are no specific restrictions to select sessions from the control group. It should be noted that, even though we select sessions using the same method as in \cite{fraser2016linguistic}, \cite{wankerl2017n}, and \cite{chen2019attention}, the number of sessions is slightly different from those works since the dataset was modified over time.

Third, we only select sessions with both the audio and the transcription information. The transcripts provide the speaker turns information. We use them to remove the clinician's speech in the data preparation step.

As a result, the data we use comprises 488 sessions (255 dementia, 233 healthy), with an average duration of $114$ seconds, recorded from 267 participants (169 dementia, 98 healthy).

We perform three preprocessing stages on the data. First, we normalize each audio signal using the average value of decibels relative to full scale (dBFS) in the data. Then, we use the speech turns information available in the dataset transcriptions to directly extract the participant's speech segments. Each segment corresponds to a participant's speech turn during the interview, thus obtaining a total of 6,267 segments (3,276 dementia, 2,991 healthy). Finally, we extend the duration of these segments by 10ms at the beginning and 10ms at the end as an attempt to mitigate speech discontinuities due to imprecisions in the turns information. In these preprocessing stages, we only use the transcripts to extract the speaker turns information, disregarding the transcript's content.

The audio files in the Pitt Corpus are single channelled (mono), sampled at a frequency of 44.1kHz and stored as PCM encoded wave files.

\subsubsection{PROMPT Database}
\label{sec:db_prompt}

%Present the PROMPT dataset (how the data was collected, how many data samples we have used, how these samples were processed, how could we separate the speech from the doctor and the patient)

The PROMPT database is part of a larger project of Keio University School of Medicine: the Project for Objective Measures Using Computational Psychiatry (PROMPT) \footnote{On the 9th of March 2016, the PROMPT project and its medical data collection were approved by the ethics committee and the Institutional Review Board of Keio University School of Medicine and by all of the other participating facilities. PROMPT protocols have been registered with the University Hospital Medical Information Network (UMIN) (UMIN ID: UMIN000023764)} \cite{kishimoto19013011}.

All the patients have given their written consent before participating in the study and, in cases in which patients were judged to be decisionally impaired, the patients' guardians provided consent. Participants were able to leave the study at any time.

In this work, we use the PROMPT Database collected from May 2, 2016 to March 31, 2019 at seven hospitals and three outpatient clinics in five different Japanese prefectures.

Speech data were recorded when the participant had free-discussion and performed several clinical tasks with trained research psychiatrists and/or psychologists. The session interviews were recorded from two synchronized microphones: one positioned close to the participant and the other placed near the clinician. The session recordings were acquired under various unconstrained acoustic conditions (i.e., with different microphones and in rooms with different reverberation characteristics).

In this work, as discussed in Section~\ref{sec:wk_dementia_assessment}, we categorize cognitive impairment based on the MMSE score, and dementia, MCI and healthy classes are defined as a MMSE score in the respective ranges of 0-23, 24-26 and 27-30. The inter-rater reliability for the MMSE score was examined and evaluated in terms of interclass correlation coefficient (ICC). The ICC assesses the consistency in the score annotations made by different raters. For the MMSE annotation in the PROMPT Database, the ICC is $0.996$ ($95\%$ CI=0.990-0.999, $p<0.01$).

The PROMPT data used in this work comprises 496 session recordings (153 dementia, 111 MCI and 232 healthy) with an average duration of 1,487 seconds, from 163 participants (49 dementia, 42 MCI and 72 healthy).

Since the PROMPT Database collects the audio recordings from two synchronized microphones, we adopted the Cross-Channel Spectral Subtraction (CCSS) method \cite{nasu2011cross} followed by a Voice Activity Detection (VAD) approach to extract the patient-only speech. The CCSS is a source separation method able to, in a meeting setting in which one microphone is prepared to each speaker, estimate the speech of a given speaker by suppressing other speakers' speech. We utilize the CCSS to suppress the doctor's speech. The resulting speech signal is then applied to a VAD, which can detect and segment the patient's speech.

% We can apply the Cross-Channel Spectral Subtraction to estimate the patient and the doctor speech and then use the power spectrum estimate at each time frame to obtain the "turns" information (or she didn't consider that and she used the silence between the patient's speech segment as well?)

After extracting the patient's speech, we apply the three preprocessing stages described in Section~\ref{sec:db_pitt} with the exception that we do not use speaker turns information to segment the patient speech.

Following the results obtained in Section~\ref{sec:exp_pitt}, we divide the patient's speech into segments of 4 seconds in a fixed-length fashion, hence resulting in 184,337 speech segments (39,593 dementia, 27,234 MCI and 117,510 healthy). In addition, we adopt a GCNN with 8 gated blocks since it resulted in the best accuracy during the experiments with the Pitt Corpus.

All the resulting audio files are single channelled (mono), sampled at the frequency of 16kHz stored as PCM encoded wave files.

\subsection{Pitt Corpus}
\label{sec:exp_pitt}

%\begin{itemize}
%\item Classification results (utterance and session level results)
%\item Experiments with different number of gated blocks
%\item Experiments with different amounts of data
%\end{itemize}

We present the average accuracy result over the 488 selected sessions of the Pitt Corpus in Table~\ref{table:comp_method}. This accuracy is computed over the session-level classification, which is obtained by applying majority voting over the speech segments' predictions. We employ SMO on the IS10 features for comparison. Our method yields the accuracy of 73.1\%, which outperforms the SMO result of 67.5\%. In Table~\ref{table:comp_method} we also report the accuracy of methods that rely on linguistic features and on the combination of linguistic and speech features, which respectively result in the best accuracies of 97.4\% and 81.9\%. Although our method has a worse performance than the linguistic features-based works presented in Table~\ref{table:comp_method}, it does not require ASR or transcription information, hence being more cost-effective and more appropriate to fast diagnosis.

\begin{table}[tb]
\centering
\small
\caption{Comparison of dementia detection methods on the Pitt Corpus interview sessions.}
\label{table:comp_method}
\begin{tabular}{lr}
\dtoprule
\textbf{Method} & \textbf{Accuracy (\%)}
\\ \dbottomrule
SMO baseline (Speech)  & 67.5
\\
Wankerl (Linguistic) \cite{wankerl2017n} & 77.1                  
\\
Fritsch (Linguistic) \cite{fritsch2019automatic} & 85.6
\\
Chen (Linguistic) \cite{chen2019attention}	& 97.4
\\
Fraser (Speech + Linguistic) \cite{fraser2016linguistic} & 81.9                   \\
\midrule[0.1pt]
Ours  & 73.1
\\ \midrule[0.2pt]
\end{tabular}
\end{table}

Table~\ref{table:confusion_matrix} shows the confusion matrix of our best model on the Pitt Corpus for the aggregated values from the ten folds both from the speech segment-level and the session-level predictions. The model is composed of eight gated blocks and it uses the speech segments obtained from the turns information in the Pitt Corpus. The confusion matrices show that predicting from a single speech segment in a session is a difficult task since the amount of information in one segment might be too limited. Thus, combining several segments for one session improves the prediction result.

\begin{table}[tb]
\centering
\footnotesize
\small
\caption{The confusion matrix depicting the classification results from all folds using the IS10 feature set as input and a GCNN with eight gated blocks on the Pitt Corpus.}
\label{table:confusion_matrix}
% \begin{tabular}{l|L{1.3cm}|R{1.3cm}|R{1.3cm}|r}
\begin{tabular}{l|L{1.3cm}|R{1.35cm}|R{1.35cm}|R{1.35cm}|} % REVIEW_X
\multicolumn{2}{c}{} & \multicolumn{3}{c}{Predicted} \\
\cline{3-5}
\multicolumn{2}{c|}{} & \multicolumn{1}{c|}{Dementia}  & \multicolumn{1}{c|}{Healthy} & \multicolumn{1}{c|}{Total}\\
\hhline{~*4{|-}|}
\multirow{2}{*}{Actual} & Dementia  & 2,340  & 936 & 3,276 \\
\hhline{~*4{|-}|} & Healthy & 1,213  & 1,778 & 2,991\\
\hhline{~*4{|-}|} & Total & 3,553  & 2,714 & Accuracy 65.7\% \\
\cline{2-5}
\multicolumn{5}{c}{}\\[-2ex]
\multicolumn{5}{c}{(a) Segment-Level Classification}\\
\multicolumn{5}{c}{}\\[-2ex]
\multicolumn{5}{c}{}\\[-2ex]
\end{tabular}

% \begin{tabular}{l|L{1.3cm}|R{1.3cm}|R{1.3cm}|r}
\begin{tabular}{l|L{1.3cm}|R{1.35cm}|R{1.35cm}|R{1.35cm}|} % REVIEW_X
\multicolumn{2}{c}{} & \multicolumn{3}{c}{Predicted} \\
\cline{3-5}
\multicolumn{2}{c|}{} & \multicolumn{1}{c|}{Dementia}  & \multicolumn{1}{c|}{Healthy} & \multicolumn{1}{c|}{Total}\\
\hhline{~*4{|-}|}
\multirow{2}{*}{Actual} & Dementia  & 189  & 66 & 255 \\
\hhline{~*4{|-}|} & Healthy & 65  & 168 & 233\\
\hhline{~*4{|-}|} & Total & 254  & 234 & Accuracy 73.1\% \\
\cline{2-5}
\multicolumn{5}{c}{}\\[-2ex]
\multicolumn{5}{c}{(b) Session-Level Classification}\\
\end{tabular}
\end{table}

\begin{figure}[t]
  \centering
  \includegraphics[width=80mm]{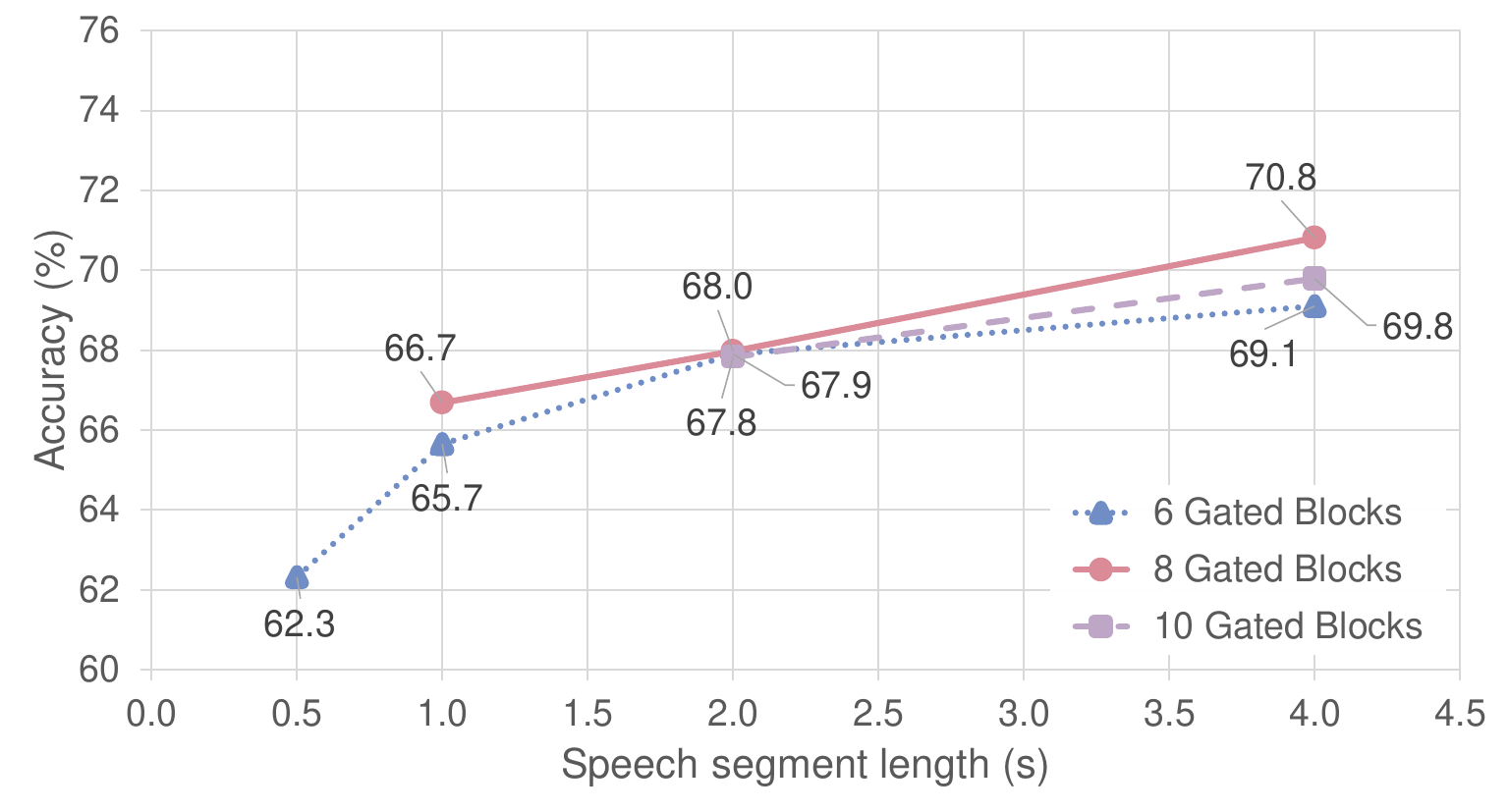} 
  \caption{Accuracy of GCNNs with different speech segment length on the Pitt Corpus}
  \label{fig:uttr_length_stat}
\end{figure}

Although our method does not depend on linguistic features, we still require the speaker turns information to segment the patient speech. In order to investigate our model's performance when the speech is partitioned into fixed-length segments and to further determine the shortest speech segment length that allows accurate data classification, we concatenate each patient's speech turns segments and we divide this concatenated speech into fixed-length segments of duration $L$. Each segment is input to the model and we apply majority voting over the segments' predictions to obtain the session-level classification. The experiment is carried out using GCNNs with a number of gated blocks equal to 6, 8 and 10 and durations $L$ of 0.5 s, 1 s, 2 s and 4 s. The results are summarized in Figure~\ref{fig:uttr_length_stat}.

%We further examine the importance of the utterance length in the classification performance on the Pitt Corpus in Figure~\ref{fig:uttr_length_stat}. We use a set of different utterance duration $L$, chosen as 0.5 s, 1 s, 2 s and 4 s. In this case, we separate each participant speech into utterances with a predetermined fixed duration $L$. Each utterance is input to the model and we apply majority voting over the utterances predictions in order to obtain the session's classification. The experiment is carried out using GCNNs with a number of gated blocks equal to 6, 8 and 10.

Figure~\ref{fig:uttr_length_stat} shows that using only 4 seconds speech segments yields results almost as good as applying the segmentation based on the speaker turns information, which suggest that there exist discriminative dementia cues in a short duration of speech data. In addition, the results in Figure~\ref{fig:uttr_length_stat} indicate that segmenting the subject's voice in the middle of their speech does not significantly degrade the performance.

\subsection{PROMPT Database}
\label{sec:exp_prompt}

We evaluate our approach on the binary classification (i.e., dementia \textit{vs} healthy) obtaining the average session-level accuracy of 80.8\%. The session-level prediction is computed as the majority voting result over all 4-seconds speech segments' predictions within a session. It should be noted that the sessions in the PROMPT Database have a longer duration compared to the Pitt Corpus, hence the accuracy is higher on the PROMPT Database.

We further evaluate our model's performance over speech intervals of different durations. Taking the 4-second fixed segment length as our duration unit, we experiment with different duration configuration for each session data, which are 4 seconds, 8 seconds, 20 seconds, 40 seconds, 1 minute, 5 minutes and all of the session speech data (i.e., the session-level prediction). In all cases, our model performs the classification over each 4-seconds speech segment. For duration configurations longer than 4 seconds, we apply majority voting to determine the prediction for each speech interval.

\begin{figure}[t]
  \centering
  \includegraphics[width=80mm]{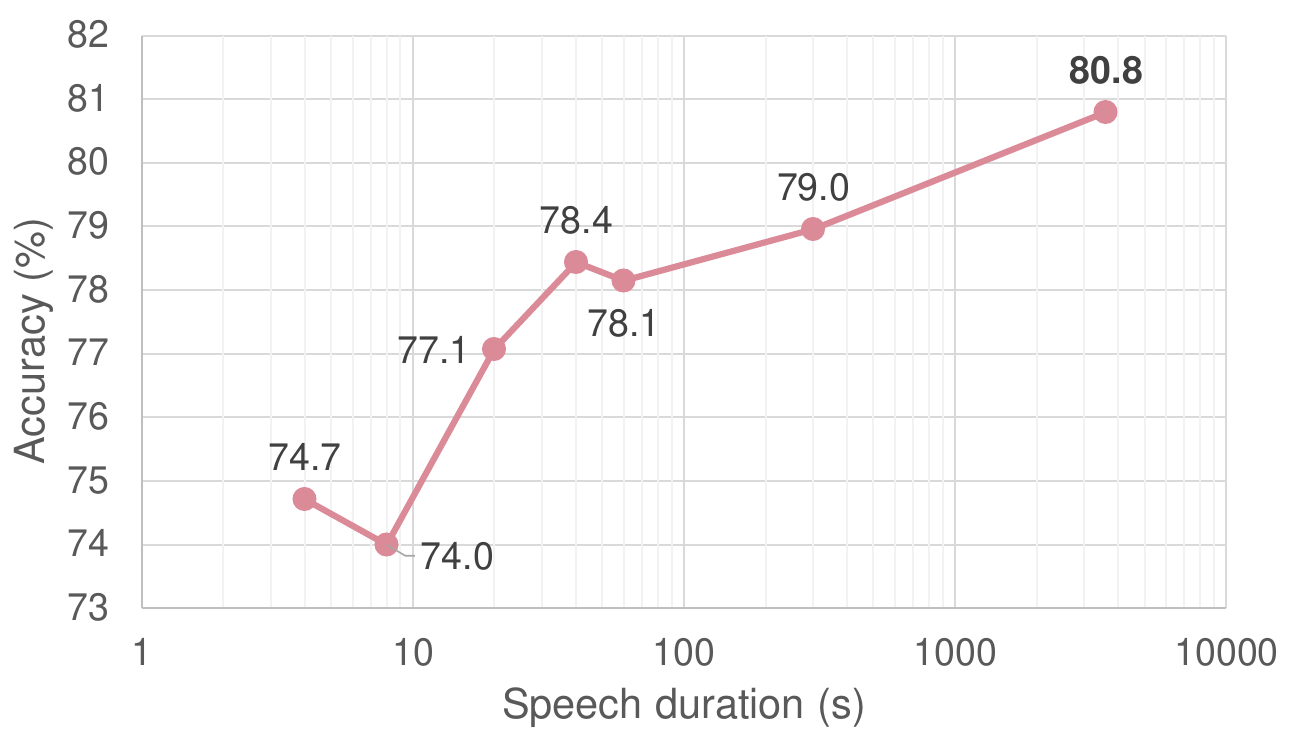}
  \caption{Accuracy of GCNN with different duration of data used for each session on the PROMPT Database. The horizontal axis is presented in logarithmic scale.}
  \label{fig:prompt_short_dur}
\end{figure}

We report our results in accuracy in Figure~\ref{fig:prompt_short_dur}. The figure shows that performance degrades if we apply shorter speech durations. However, we obtain the average accuracy of 77.1\% by using only 20 seconds of data for each session and the average accuracy of 74.7\% when we use only 4 seconds of data. Although there is still room for improvement, this result represents an important step towards the application of automatic dementia detection tools to real-world diagnosis, in which, every so often, there is little available speech data. We have additionally reported the segment-level classification confusion matrix for the 4-second segmentation of one of our folds in Table~\ref{table:confusion_matrix-prompt}.

\begin{table}[tb]
\centering
\footnotesize
\small
\caption{Segment-level confusion matrix of one of the folds on the PROMPT database for the classification based on 4 seconds long speech segments.}
\label{table:confusion_matrix-prompt}
% \begin{tabular}{l|L{1.3cm}|R{1.3cm}|R{1.3cm}|r}
\begin{tabular}{l|L{1.3cm}|R{1.35cm}|R{1.35cm}|R{1.35cm}|} % REVIEW_X
\multicolumn{2}{c}{} & \multicolumn{3}{c}{Predicted} \\
\cline{3-5}
\multicolumn{2}{c|}{} & \multicolumn{1}{c|}{Dementia}  & \multicolumn{1}{c|}{Healthy} & \multicolumn{1}{c|}{Total}\\
\hhline{~*4{|-}|}
\multirow{2}{*}{Actual} & Dementia  & 3,882  & 380 & 4,262 \\
\hhline{~*4{|-}|} & Healthy & 3,227  & 7,418 & 10,645\\
\hhline{~*4{|-}|} & Total & 7,109 & 7,798 & Accuracy 75.80\% \\
\cline{2-5}
\multicolumn{5}{c}{}\\[-2ex]
\end{tabular}
\end{table}

% Comparable sensitivity and specificity is achieved with the best result of 83.7\% and 89.5\% respectively for 5 minutes of patient speech. 
% We report our results in accuracy, sensitivity, and specificity as can be seen in Figure~\ref{fig:prompt_short_dur} for a more comprehensive evaluation.

%\begin{table}[]
%\centering
%\footnotesize
%\small
%  \caption{Confusion matrix of the evaluation on PROMPT Database of Condition D \texttt{vs} M \texttt{vs} H.}
%  \label{table:prompt_confusion_matrix}
%\begin{tabular}{l|L{0.7cm}|R{0.7cm}|R{0.7cm}|R{0.7cm}|r}
%\multicolumn{2}{c}{}                                  & \multicolumn{3}{c}{Predicted}   & \\
%\cline{3-5}
%\multicolumn{2}{c|}{}                      & \multicolumn{1}{c|}{D}        & \multicolumn{1}{c|}{M}       & \multicolumn{1}{c|}{H}       & \\
%\hhline{~*4{|-}|}
%\multirow{2}{*}{Actual}          & D       & 131    & 65   & 26  & \\
%\hhline{~*4{|-}|}                & M       & 12     & 16   & 15   & \\
%\hhline{~*4{|-}|}                & H       & 18     & 16   & 157   & \\
%\cline{2-5} 
%\multicolumn{5}{c}{}\\[-2ex]
%\end{tabular}
%\end{table}

We further perform a more comprehensive experiment on the binary classification with various configurations. Apart from distinguishing dementia and healthy sessions, we perform the classification on the dementia versus non-dementia case and the healthy versus non-healthy case by respectively adding the participants with a MMSE score in the MCI range to the healthy and to the dementia groups. Table~\ref{table:prompt_exp_2class} depicts these configurations in the column ``Conditions'' and it reports the results in terms of accuracy for these configurations and different speech durations.

\begin{table*}[tb]
\centering
\small
\caption{Experimental results on the PROMPT Database for different experiment conditions for the binary classification. The dementia, MCI and healthy classes are respectively represented by the characters `D', `M' and `H'.}
\label{table:prompt_exp_2class}
\begin{tabular}{@{}lrrrrrrr@{}}
\dtoprule
\multicolumn{1}{c}{\multirow{2}{*}{\textbf{Conditions}}} & \multicolumn{7}{c}{\textbf{Accuracy (\%)}}                                                                                                  \\ \cmidrule(l){2-8}
\multicolumn{1}{c}{} & \multicolumn{1}{c}{\textbf{4 sec}}             & \multicolumn{1}{c}{\textbf{8 sec}}             & \multicolumn{1}{c}{\textbf{20 sec}}      & \multicolumn{1}{c}{\textbf{40 sec}}            & \multicolumn{1}{c}{\textbf{1 min}}        & \multicolumn{1}{c}{\textbf{5 min}}       & \multicolumn{1}{c}{\textbf{all data}}    \\ \dbottomrule
D \textit{vs} M + H            & 73.8           & 74.3           & 77.9     & \textbf{78.6}  & 78.3      & 77.0              & 77.6          \\
D + M \textit{vs} H            & 71.3           & 70.9           & 73.6     & 74.0           & 74.1      & \textbf{75.9}     & 74.3          \\
D \textit{vs} H                & 74.7           & 74.0           & 77.1     & 78.4           & 78.1      & 79.0              & \textbf{80.8} \\ \midrule[0.2pt]

\end{tabular}
\end{table*}

%\\
%\dtoprule
%\multicolumn{1}{c}{\multirow{2}{*}{\textbf{Conditions}}} & \multicolumn{7}{c}{\textbf{F1 Score (\%)}}                                                                                                     \\ \cmidrule(l){2-8}
%\multicolumn{1}{c}{}  & \multicolumn{1}{c}{\textbf{4 sec}}             & \multicolumn{1}{c}{\textbf{8 sec}}             & \multicolumn{1}{c}{\textbf{20 sec}}      & \multicolumn{1}{c}{\textbf{40 sec}}            & \multicolumn{1}{c}{\textbf{1 min}}        & \multicolumn{1}{c}{\textbf{5 min}}       & \multicolumn{1}{c}{\textbf{all data}}    \\ \dbottomrule
%D \textit{vs} M + H  & 73.2 & 73.1 & 77.4 & \textbf{78.0} & 77.8 & 76.4 & 77.0 \\
%D + M \textit{vs} H  & 70.9 & 70.4 & 73.1 & 73.5 & 74.0 & \textbf{75.5} & 73.9 \\
%D \textit{vs} H & 74.5 & 73.3 & 76.8 & 78.0 & 77.9 & 78.7 & \textbf{80.5}  \\ \midrule[0.2pt]

In terms of session classification accuracy, adding the MCI class yields worse performance on both Condition D \textit{vs} M + H (78.6\%) and Condition D + M \textit{vs} H (75.9\%) compared to Condition D \textit{vs} H (80.8\%) as it can be seen in Table~\ref{table:prompt_exp_2class}. Moreover, it is possible to observe that, for the experiments that include the MCI class data, the accuracy does not increase monotonously with the amount of speech data. This result suggests that we should not combine the MCI participants either with healthy or dementia participants. This might be explained from the fact that MCI participants cannot be considered healthy due to their cognitive ability decline, but MCI cannot be framed as dementia either, since this decline is less severe compared to dementia. It is also interesting to see that combining MCI with dementia patients yields worse performance than combining MCI with healthy subjects. Further investigation on the closer relation between MCI patients and healthy subjects might be needed based on this result.

Finally, we also conduct the three-class classification to distinguish between sessions with dementia, MCI and healthy subjects. The results are reported in Table~\ref{table:prompt_exp_3class}. We obtain the average accuracy of 65.0\% using 4 seconds of session data. While MCI patients might present subtle different visible characteristic from healthy or dementia patients, they are very different in actual. Detecting MCI patients is important for the early prediction of dementia but it is difficult due to the nearly ambiguous nature of the data and the lack of a medical standard to classify MCI patients, as discussed in Section~\ref{sec:wk_dementia_assessment}.

\begin{table*}[tb]
\centering
% \footnotesize
\scriptsize % REVIEW
\small
\caption{Experimental results on the PROMPT Database for the three-class classification of D \textit{vs} M \textit{vs} H. The results are reported in accuracy (\%).}
\label{table:prompt_exp_3class}
\begin{tabular}{@{}rrrrrrr@{}}
\dtoprule
\multicolumn{1}{c}{\textbf{4 sec}} & \multicolumn{1}{c}{\textbf{8 sec}} & \multicolumn{1}{c}{\textbf{20 sec}} & \multicolumn{1}{c}{\textbf{40 sec}} & \multicolumn{1}{c}{\textbf{1 min}}   & \multicolumn{1}{c}{\textbf{5 min}}  & \multicolumn{1}{c}{\textbf{all data}}    \\ \dbottomrule
65.0          & \textbf{61.1}  & 58.4     & 60.6           & 59.2      & 57.7           & 58.3          \\
\midrule[0.2pt]
\end{tabular}
\end{table*}
%
%F1 Score (\%)  & 47.4 & 48.3  & 49.4 & \textbf{51.6} & 51.2 & 50.2 & 51.4          \\

\section{Conclusion}
\label{conclusion}
%conclusion here. Future work: MCI

We present a method for dementia detection solely based on speech data. Using a GCNN architecture on top of the IS10 paralinguistic feature set yields the best accuracy of 73.1\% in an English dataset, the Pitt Corpus, and 80.8\% in a Japanese dataset, the PROMPT Database. We achieve the accuracy of 77.1\% by using only 20 seconds of data on the PROMPT Database and 74.7\% when we consider only 4 seconds of data. These results show our model's capability of making predictions with a reduced amount of data, which is important for real-world dementia diagnosis. We further perform the three-class classification of dementia, MCI and healthy subjects on the PROMPT Database, which yields the accuracy of 60.6\%.

Even though our results on the Pitt Corpus are worse when compared to the current linguistic approaches, our method is cost-effective since it does not require any transcription data and it allows the detection result to be obtained faster, which is particular promising to the early diagnosis of dementia. Moreover, our method may be applicable to resource-deficient language speakers more easily than methods that rely on linguistic information. This is because it is difficult to build a language model and a high-accuracy ASR for those languages.

Nevertheless, there are still remaining improvements to enable our model to perform diagnosis in real case scenarios. With that said, in the near future, we intend to analyse the temporal pattern of dementia patients and to incorporate more modalities (e.g., facial features, body motion). Moreover, we would like to analyse the similarities and the differences of the MCI patients' data to the other classes in order to improve the detection of MCI patients.

\section*{Acknowledgements}
\label{acknowledgements}

This work was supported by JST CREST {[grant numbers JPMJCR1687, JPMJCR19F5]}; JSPS KAKEN {[grant number 16H02845]}, and the Japan Agency for Medical Research and Development (AMED) {[grant number JP18he1102004]}.

\bibliographystyle{ieicetr} % bib style
\bibliography{bibliography} % your bib database

\end{document}